\documentclass[pra,twocolumn,preprintnumbers,amsmath,amssymb,nofootinbib]{revtex4}
\usepackage{amsmath}
\usepackage{amssymb}
\usepackage{slashed}
\usepackage{graphicx}
\usepackage{graphics}
\usepackage{epsfig}
\usepackage{psfrag}
\usepackage{hyperref}

\usepackage{color}

\newcommand{\E}{\mathrm{e}}
\newcommand{\I}{\mathrm{i}}
\newcommand{\be}{\begin{eqnarray}}
\newcommand{\ee}{\end{eqnarray}}

\newcommand{\yb}{\bar\psi}

\newcommand{\Nf}{N_{\text{f}}}

\newcommand{\pat}{\partial_t}
\newcommand{\Eqref}[1]{Eq.~\eqref{#1}}

\begin{document}

\title{Asymptotic safety of simple Yukawa systems}
%\date{\today}
\author{Holger Gies${}^1$ and Michael M. Scherer${}^{1,2}$}

\pacs{}

\affiliation{\mbox{\it ${}^1$Theoretisch-Physikalisches Institut, Friedrich-Schiller-Universit{\"a}t Jena,
Max-Wien-Platz 1, D-07743 Jena, Germany}
\mbox{\it ${}^2$Institut f{\"u}r Theoretische Physik, Universit{\"a}t Heidelberg,
Philosophenweg 16, D-69120 Heidelberg, Germany} 
\mbox{\it E-mail: { holger.gies@uni-jena.de, michael.scherer@uni-jena.de}}
}

\begin{abstract} 
  We study the triviality and hierarchy problem of a $Z_2$-invariant Yukawa
  system with massless fermions and a real scalar field, serving as a toy
  model for the standard-model Higgs sector. Using the functional RG, we look
  for UV stable fixed points which could render the system asymptotically
  safe. Whether a balancing of fermionic and bosonic contributions in the RG
  flow induces such a fixed point depends on the algebraic
  structure and the degrees of freedom of the system.  Within the region of
  parameter space which can be controlled by a nonperturbative next-to-leading
  order derivative expansion of the effective action, we find no non-Gau\ss
  ian fixed point in the case of one or more fermion flavors. The
  fermion-boson balancing can still be demonstrated within a model system with
  a small fractional flavor number in the symmetry-broken regime. The UV
  behavior of this small-$\Nf$ system is controlled by a conformal Higgs
  expectation value. The system has only two physical parameters, implying
  that the Higgs mass can be predicted. It also naturally explains the heavy
  mass of the top quark, since there are no RG trajectories connecting the UV
  fixed point with light top masses. 
\end{abstract}

\maketitle

\section{Introduction and Summary}

\subsection{Triviality and Hierarchy in the standard model}

The standard model of particle physics is a successful theory that has passed
a number of high-precision tests. One crucial building block is the Higgs sector
which renders the perturbative expansion of correlation functions well defined
and parameterizes the masses of matter fields and weak gauge bosons. So far,
the Higgs sector has been indirectly tested by the precision data but will be
directly explored at the LHC. Beyond this success, two problems of the
standard model have been particularly inspiring to design new ideas for a
larger fundamental framework: the hierarchy problem and the triviality
problem. 

Whereas the hierarchy problem is not a fundamental problem in the sense of
rendering the standard model ill-defined, the triviality problem truly
inhibits an extension of the standard model to arbitrarily high momentum
scales. Loosely speaking, the scale of maximum ultra-violet (UV) extension
$\Lambda_{\text{UV,max}}$ induced by triviality is related to the Landau pole
of perturbation theory.  The standard model has triviality problems in both
the Higgs sector
\cite{Wilson:1973jj,Luscher:1987ek,Hasenfratz:1987eh,Heller:1992js,%
Callaway:1988ya,Rosten:2008ts} as well as the U(1) gauge sector
\cite{Landau,Gell-Mann:fq,Gockeler:1997dn,Gies:2004hy}. But since
$\Lambda_{\text{UV, max}}$ of the Higgs sector is much smaller than that of
the U(1) sector \cite{Gies:2004hy}, evading triviality in the Higgs sector is
of primary importance. In fact for heavy Higgs boson masses, the scale of
maximum UV extension $\Lambda_{\text{UV,max}}$ could be much smaller than the
Planck or GUT scale
\cite{Cabibbo:1979ay,Kuti:1987nr,Hambye:1996wb,Fodor:2007fn,Gerhold:2008mb}. For
supersymmetrically extended models, $\Lambda_{\text{UV,max}}$ can even be
smaller than in the standard model \cite{Yndurain:1991vm}.

Traces of the triviality problem can already be found in perturbation
theory. The relation between bare and renormalized four-Higgs-boson coupling
$\lambda \phi^4$ in one-loop RG-improved perturbation theory is given by
\begin{equation}
 \frac{1}{\lambda_{\text R}}-\frac{1}{\lambda_{\Lambda}}=\beta_0\ \mbox{Log}\left(
   \frac{\Lambda}{m_{\text R}}\right),\ \beta_0=\text{const.}>0,\label{eq:betarun} 
\end{equation}
where $\lambda_\Lambda$ is the bare and $\lambda_{\text R}$ the renormalized
coupling; $\Lambda$ is the UV cutoff scale and $m_{\text R}$ denotes a
renormalized mass scale. The first $\beta$ function coefficient $\beta_0$ is
generically positive for $\phi^4$ theories. Keeping $\lambda_{\text R}$ and
$m_{\text R}$ fixed, say as measured in an experiment at an infrared (IR)
scale, an increase of the UV cutoff $\Lambda$ has to be compensated by an
increase of the bare coupling $\lambda_\Lambda$. But $\lambda_\Lambda$
eventually hits infinity at a finite UV scale $\Lambda_{\text{L}}=m_{\text
  R}\, \exp [1/(\beta_0 \lambda_{\text R})]$. This Landau pole provides
for a first estimate of the scale of maximum UV extension $\Lambda_{\text
  L}\simeq \Lambda_{\text{UV,max}}$.

Of course, perturbation theory is useless for a reliable estimate of
$\Lambda_{\text{UV,max}}$, since it is an expansion about zero coupling. Near
the Landau pole, nonperturbative physics can set in and severely modify the
picture. In fact, the QED Landau pole has been shown to be outside the
physical parameter space, since it is screened by the nonperturbative
phenomenon of chiral symmetry breaking
\cite{Gockeler:1997dn,Gies:2004hy}. (Still, there remains a finite scale of
maximum UV extension $\Lambda_{\text{UV,max}}<\infty$.) A study of the
triviality problem therefore mandatorily requires a nonperturbative tool.

The hierarchy problem of the standard model is a fine-tuning problem of
initial conditions and arises also from the properties of the Higgs
sector. For generic values of the initial squared bare Higgs mass scale
$m_\Lambda^2\sim \Lambda^2$ at UV cutoff $\Lambda$, the system is either in
the symmetric phase exhibiting no electroweak symmetry breaking, or it is in
the broken phase with gauge, Higgs-boson and fermion masses of order
$\Lambda$. Both phases are separated by a quantum phase transition at a
critical value $m_{\Lambda,\text{cr}}^2$. A large hierarchy, i.e., a large
separation of particle masses from the cutoff scale, requires an extremely
fine-tuned value of $m_{\Lambda}^2$ close to the critical value. For instance,
separating the scale of electroweak symmetry breaking $\Lambda_{\text{EW}}$
from the UV scale, e.g., given by a GUT scale $\Lambda_{\text{GUT}}$, requires
to fine-tune $m_{\Lambda}^2$ to $m_{\Lambda, \text{cr}}$ within a precision of
$\Lambda_{\text{EW}}^2/\Lambda_{\text{GUT}}^2\sim 10^{-28}$.

This problem of ``unnatural'' initial conditions stems from the fact that the
mass parameter in the Higgs sector renormalizes quadratically $\sim
\Lambda^2$. In a renormalization group (RG) language, the critical value
$m_{\Lambda,\text{cr}}^2$ denotes a strongly IR repulsive fixed point with
critical exponent $\Theta=2$. Again, these statements hold in the vicinity of
the Gau\ss ian fixed point, where all couplings are small and perturbation
theory holds. We stress that there is nothing conceptually wrong with
fine-tuned initial conditions, but it is generally considered as
unsatisfactory. A solution of the hierarchy problem should either explain the
fine-tuned initial conditions within an underlying theory or correspond to a
UV extension of the standard model which has no critical exponents
significantly larger than zero, implying a slow, say logarithmic, running of
the parameters.

\subsection{Asymptotically Safe Scenarios}

This work is devoted to an investigation of a simple Yukawa model, serving as
a toy-model for the Higgs sector of the standard model. Using the functional
RG as a nonperturbative approach, we systematically search for the existence
of non-Gau\ss ian interacting fixed points of the RG flow which could
facilitate an extension of the model to arbitrarily high momentum scales. If
such a system sits on a trajectory that hits the fixed point in the UV, the UV
cutoff can safely be taken to infinity, $\Lambda\to \infty$, and the theory is
{\em asymptotically safe} in Weinberg's sense \cite{Weinberg:1976xy}, see
\cite{Percacci:2007sz} for a recent review. Such fixed points would not only
solve the triviality problem by construction, but could have small critical
exponents as required for a solution of the hierarchy problem. 

A paradigm example for an asymptotically safe theory are four-fermion models
such as the Gross-Neveu model in $2< D< 4$ dimensions
\cite{Rosenstein:pt}. Even though these models are perturbatively not
renormalizable and thus seemingly trivial, they are nonperturbatively
renormalizable at a non-Gau\ss ian fixed point and hence can be extended to
arbitrarily high scales. A similar result has been found for nonlinear sigma
models in $d>2$ \cite{Codello:2008qq}. Recently, even the program of
quantizing gravity with the aid of a non-Gau\ss ian fixed point within an
asymptotic-safety scenario has become exceedingly successful
\cite{Reuter:1996cp}; this also includes certain models with a nontrivial
scalar sector \cite{Percacci:2003jz}. Asymptotically safe scenarios have also
been successfully developed for extra-dimensional gauge theories
\cite{Gies:2003ic}. Furthermore standard-model alternatives without a
fundamental Higgs field have been based on asymptotically safe fermionic
scenarios \cite{Gies:2003dp}, which, however, are generically plagued by a
strong hierarchy problem; see also \cite{Schwindt:2008gj} for an
asymptotically free fermionic model where nonlocal interactions lead to an
improved hierarchy.

In the context of the Higgs sector of the standard model, an asymptotically 
safe scenario can have further advantages in addition to solving the fundamental problems of triviality and hierarchy.  Contrary to many other
ideas for addressing these problems, our approach does not introduce new
degrees of freedom or further symmmetries and thus no new parameters. Indeed,
asymptotic safety can lead to a reduction of parameters and hence have more
predictive power, since the number of physical parameters to be determined by
experiment depends on the properties of the non-Gau\ss ian fixed point, i.e.,
on the dynamics of the theory.  The price to be paid is that we have to give
up on the requirement of perturbativity \cite{Cabibbo:1979ay}, stating that
the standard model should never leave the Gau\ss ian fixed-point regime, being
the domain of validity of perturbative theory. There might indeed be no
fundamental reason for this requirement except for saving human beings from
having to deal with nonperturbative problems.

A crucial question for all nonperturbative techniques is the systematic
consistency and reliability of the results. In this work, we compute the RG
flow of the model in a systematic derivative expansion of the effective
action. This expansion is controlled if the momentum dependence of full
effective vertices takes only little influence on the flow. A direct means for
measuring this influence is the size of the anomalous dimensions $\eta$ of the
fields, since next-to-leading order contributions couple to the leading-order
derivative expansion only via terms $\sim\eta$. Monitoring the size of $\eta$
thus gives us a direct estimate of the reliability of our results. 

\subsection{Asymptotic Safety of Yukawa Systems}

Our results can be summarized in a simple picture: the RG flow of the Yukawa
model can be in the symmetric (SYM) regime or in the regime of spontaneous
symmetry breaking (SSB) where the bosonic field expectation value $v$ is
nonzero, $v>0$. Whereas we do not find interacting fixed points in the SYM
regime, the structure of the flow becomes richer in the SSB regime, since new
interactions can be mediated by the condensate. However, the flow in the SSB
regime is typically characterized by a {\em freeze-out} of all couplings. This
is because all particles coupling to the vacuum expectation value (vev)
acquire a mass and decouple from the flow. Our central idea is that the
contributions with opposite sign from bosonic and fermionic fluctuations to
the vev can be balanced such that the vev exhibits a conformal behavior, $v\sim
k$. Here, $k$ is the scale at which we consider the couplings of the
system. The dimensionless squared vev $\kappa= \frac{1}{2}v^2/k^2$ has a flow equation of
the form
\begin{equation}
\pat \kappa \equiv \pat \frac{v^2}{2 k^2} = -2 \kappa + \text{interaction terms},
\quad \pat = k\frac{d}{dk}.
\end{equation}
If the interaction terms are absent, the Gau\ss ian fixed point $\kappa=0$ is
the only conformal point, corresponding to a free massless theory. If the
interaction terms are nonzero, e.g., if the couplings approach interacting
fixed points by themselves, the sign of these terms decides about a possible
conformal behavior. A positive contribution from the interaction terms gives rise to a fixed point at
$\kappa>0$ which can control the conformal running over many scales. If
they are negative, no conformal vev is possible. Since fermions and bosons
contribute with opposite signs to the interaction terms, the existence of a
fixed point $\kappa_\ast>0$ crucially depends on the relative strength between
bosonic and fermionic fluctuations.  More specifically, the bosons have to win
out over the fermions.

\begin{figure}[ht]
\centering
\includegraphics[width=0.47\textwidth]{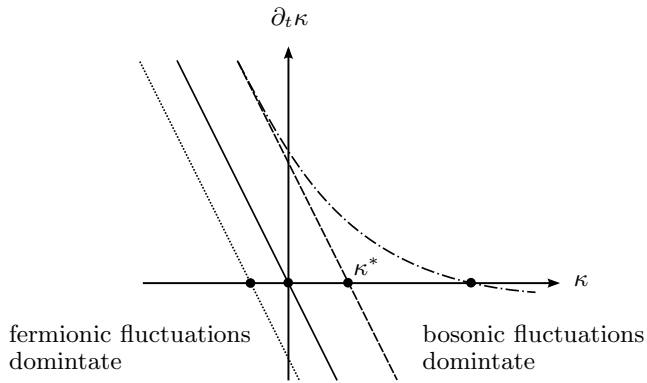}
\caption{Sketch of the flow of the dimensionless squared Higgs vacuum
  expectation value $\kappa$. Solid line: free massless theory,
  $\pat\kappa=-2\kappa$, with a trivial Gau\ss ian fixed point at
  $\kappa=0$. Dotted line: fermions dominate, inhibiting an acceptable
  non-Gau\ss ian fixed point for positive $\kappa$.  Dashed line: bosonic
  fluctuations dominate, inducing a non-Gau\ss ian fixed point $\kappa^\ast>0$
  with conformal behavior of the Higgs expectation value. In the present set
  of Yukawa systems, this behavior occurs at small $\Nf$, $\Nf \lesssim
  0.3$. Dot-dashed line: interaction-induced reduction of the slope at the
  fixed point (critical exponent), implying an improvement of the hierarchy
  behavior.} 
\label{fig:sketch}
\end{figure}

In Fig.~\ref{fig:sketch}, we sketch various options for the flow of the
dimensionless squared vev $\kappa$. The solid line depicts the free massless
theory with a trivial Gau\ss ian fixed point at $\kappa=0$ . If the fermions
dominate, the interaction terms are negative and the fixed point is shifted to
negative values (being irrelevant for physics), cf. dotted line. If the
bosonic fluctuations dominate, the $\kappa$ flow develops a non-Gau\ss ian
fixed point at positive values $\kappa^\ast>0$ that can support a conformal
behavior over many orders of magnitude, cf. dashed line. This fixed point is
UV attractive, implying that the vev is a relevant operator near the fixed
point. If the interaction terms are approximately $\kappa$ independent, the
slope of $\pat\kappa$ near the fixed point is still close to $-2$,
corresponding to a critical exponent $\Theta\simeq 2$ and a persistent
hierarchy problem. An improvement of ``naturalness'' could arise from a
suitable $\kappa$ dependence of the interaction terms that results in a
flattening of the $\kappa$ flow near the fixed point, cf. dot-dashed
line. Whether or not this happens is a result of and can be computed within
the theory.

For the simple Yukawa systems considered here, we find fermionic dominance for
one or more fermion flavors, excluding an asymptotic safety scenario of this
kind. In fact, a careful analysis of the model in the SYM regime as well as the SSB
regime reveals that no non-Gau\ss ian fixed point exists in the accessible
parameter range neither for strong coupling nor induced by balanced threshold
behavior. Nevertheless, this fermion dominance is not an unavoidable property
of the system, but a result of the algebraic details of the model. This is
illustrated by treating the flavor number $\Nf$ as a continuous variable. We
observe boson dominance for $\Nf\lesssim 0.3$, implying the existence of a
suitable non-Gau\ss ian fixed point and a non-trivial interacting fundamental
theory valid on all scales. We also find a dependence of the largest critical
exponent on the flavor number, supporting also reductions from the
value $\Theta=2$. However, a significant hierarchy always remains in the
simple models considered here.

Two features render our mechanism particularly attractive: first, the UV
fixed-point properties are such that the system has one parameter less than
expected; in other words, once the top mass and the Higgs vev are fixed, the
Higgs mass is a true prediction of the theory. Second, the set of all possible
flows that start from the UV fixed-point lead to a constrained set of physical
low-energy values; most importantly, the top mass does generically not become
much lighter than the vev, offering a natural explanation for the large mass
of the top.

This paper is organized as follows: In Sect. \ref{sec:RG}, we briefly recall
the functional RG, review Weinberg's asymptotic-safety scenario, and present
the model and our nonperturbative approximation techniques. In
Sect. \ref{sec:SYM}, we analyse the model in the symmetric regime, whereas
Sect. \ref{sec:SSB} is devoted to a study of the regime of spontaneously
broken symmetry for integer flavor number $\Nf=1$. The possibility of an
alternative asymptotically free scenario based on Halpern-Huang scalar
potentials is briefly investigated in Sect. \ref{sec:HH}. The asymptotically
safe scenarios at small flavor number $\Nf$ are presented in
Sect. \ref{sec:Nf}, and their predictive power is illustrated in
Sect. \ref{sec:PP}. Conclusions are presented in Sect. \ref{sec:conc}.

\section{Renormalization flow of simple Yukawa systems}
\label{sec:RG}

\subsection{Flow equation and Asymptotic safety}

The search for asymptotically safe field theories adds an interesting twist to
the realm of quantum field theoretic problems. Standard problems start from a
microscopic bare action that has to be quantized in order to construct quantum
correlation functions. In the present case, we do not even know the
microscopic action to be quantized, but we try to construct it by searching
for a non-Gau\ss ian fixed point action in the space of all possible action
functionals, i.e., in {\em theory space}. In other words, we wish to quantize
a generic action of a system which is specified by its degrees of freedom and
its symmetries.

This is indeed possible with the aid of the functional RG which provides a
vector field $\boldsymbol \beta$ in theory space in terms of RG $\beta$-functions 
for all possible generalized couplings $(\boldsymbol\beta)_i =
\beta_{g_i}(g_1,g_2,\dots)\equiv \pat g_i$. The functional RG can conveniently be formulated in
terms of a flow equation for the effective average action $\Gamma_k$, the
Wetterich equation \cite{Wetterich:1993yh}:
\begin{equation}\label{flowequation}
	\partial_t\Gamma_k[\Phi]
        =\frac{1}{2}\mathrm{STr}\{[\Gamma^{(2)}_k[\Phi]+R_k]^{-1}(\partial_tR_k)\},
        \quad \pat=k\frac{d}{dk}  
	.
\end{equation}
Here, $\Gamma^{(2)}_k$ is the second functional derivative with respect to the
field $\Phi$ (the latter representing a collective field variable for all
bosonic or fermionic degrees of freedom), and $R_k$ denotes a
momentum-dependent regulator function that suppresses IR modes below a
momentum scale $k$. The solution to the Wetterich equation provides for an RG
trajectory in theory space, interpolating between the bare action $S_\Lambda$ to be
quantized $\Gamma_{k\to\Lambda}\to S_\Lambda$ and the full quantum effective
action $\Gamma=\Gamma_{k\to 0}$, being the generating functional of 1PI
correlation functions; for reviews, see \cite{ReviewRG}. 

A quantum field theory can be considered as fundamental, if its RG trajectory
can be extended over all scales which
requires that the UV cutoff $\Lambda$ can be sent to infinity. This is, for
instance, possible if the trajectory approaches a fixed point in theory
space. Let us assume that the effective average action $\Gamma_k$ can be
parameterized by a possibly infinite set of generalized dimensionless
couplings $g_i$. Then, the Wetterich equation provides us with the flow of
these couplings  $\pat g_i= \beta_{g_i}(g_1,g_2,\dots) $. A fixed point
$g_i^\ast$ is defined by
\begin{equation}
 \beta_i(g_1^{\ast},g_2^{\ast},...)=0\ \forall \ i\,.
\end{equation}
The fixed point is non-Gau\ss ian, if at least one fixed-point coupling is
nonzero $g_j^\ast\neq 0$. If the RG trajectory flows into a fixed point in the
UV, the UV cutoff can safely be taken to infinity and the system approaches a
conformally invariant state for $k\to\infty$. 

In addition to being fundamental, we also want the theory to be
predictive. For this, we consider the fixed-point regime, where the flow can
be linearized,
\begin{equation}
 \partial_t g_i = B_i{}^j (g^\ast_j-g_j)+\dots, \quad B_i{}^j =\frac{\partial
   \beta_{g_i}}{\partial g_j} \Big|_{g=g^\ast}.\label{eq:lin}
\end{equation}
Diagonalizing the stability matrix $B_i{}^j$,
\begin{equation}
B_i{}^j\, V_j^I =-\Theta^I V_i^I,
\end{equation}
in terms of right-eigenvectors $V_i^I$ which are enumerated by the index $I$,
the resulting {\em critical exponents} $\Theta^I$ allow for a classification
of physical parameters. The solution of the coupling flow in the fixed-point
regime is given by
\begin{equation}
g_i=g_i^\ast + \sum_I C^I\, V_i^I \, \left(\frac{k_0}{k} \right)^{\Theta^I},
\end{equation}
where the integration parameters $C^I=$const. define the initial conditions at
a reference scale $k_0$. Whereas all eigendirections with $\Theta^I<0$ die out
towards the IR and thus are irrelevant, all relevant directions with $\Theta^I>0$
increase towards the IR and thus determine the macroscopic physics (for the
marginal directions $\Theta^I=0$, it depends on the higher-order terms in the
expansion about the fixed point). Hence the number of relevant and
marginally-relevant directions determines the number of physical parameters to
be fixed. The theory is predictive if this number is finite. 

If a critical exponent is much larger than zero, say of $\mathcal O(1)$, the
RG trajectory rapidly leaves the fixed-point regime towards the IR. Therefore,
separating a typical UV scale where the system is close to the fixed point
from the IR scales where, e.g., physical masses are generated requires a
significant fine-tuning of the initial conditions. In the context of the
standard model,  the size of the largest $\Theta^I$ is a quantitative measure
of the hierarchy problem. 

For the flow towards the IR, the linearized fixed-point flow \Eqref{eq:lin}
generally is insufficient and the full nonlinear $\beta$ functions have to be
taken into account. Even the parameterization of the effective action in terms
of the same degrees of freedom in the UV and IR might be
inappropriate. Nevertheless, we use the same bosonic and fermionic degrees of
freedom on all scales in the present work, since we specifically want to
address the question whether standard-model IR degrees of freedom can have an
interacting UV completion.

\subsection{Yukawa system}

In the present work, we concentrate on a $Z_2$ invariant Yukawa theory,
involving one real scalar field and $N_f$ Dirac fermions. This model mimics
the Higgs sector of the standard model, in the sense that spontaneous symmetry
breaking (SSB) in the scalar sector can generate fermion masses. Also, since
the symmetry is discrete, no Goldstone bosons are generated in the SSB
phase.

In the spirit of perturbative power-counting near the Gau\ss ian fixed point,
the microscopic action subject to perturbative quantization would read
\begin{equation}
  S = \!\int\!\! d^4{x} \left(
    \frac{1}{2}(\partial_{\mu}\phi)^2+\frac{\bar{m}^2}{2}\phi^2
+\frac{\bar{\lambda}}{8}\phi^4+\bar{\psi}\mbox{i}\slashed{\partial}\psi
+\mbox{i}\bar{h}\phi\bar{\psi}\psi\right),  
\end{equation}
where the bare parameter space would be spanned by the boson mass $\bar m$,
the boson interaction $\bar\lambda$ and the Yukawa coupling $\bar h$. 

In order to allow for more general actions also near interacting fixed points,
we only impose that the action be symmetric under the discrete chiral symmetry,
\begin{equation}
\psi\to \E^{\I \frac{\pi}{2} \gamma_5} \psi, \quad
\yb\to \E^{\I \frac{\pi}{2} \gamma_5} \yb, \quad
\phi\to -\phi.\label{eq:discsym}
\end{equation}
Furthermore, we restrict the action to bilinears in the fermions and expand
the bosonic part in powers of field derivatives. To next-to-leading order, our
truncation of the effective action thus reads
\begin{equation}\label{ansatz}
	\Gamma_k \!= \!\!\int\!\! d^4x \!\left(
          \frac{Z_{\phi,k}}{2}(\partial_{\mu}\phi)^2+U_k(\rho)+Z_{\psi,k}\bar{\psi}
          \mbox{i}\slashed{\partial}\psi+\mbox{i}\bar{h}_k\phi\bar{\psi}\psi\right)\!, 
\end{equation}
where $\rho=\frac{1}{2}\phi^2$. We also confine ourselves to a Yukawa term
linear in $\phi$. In many cases, it suffices qualitatively to expand the
effective potential in powers of $\rho$, see below. Keeping the wave function
renormalizations $Z_{\phi,k},Z_{\psi,k}$ fixed defines the leading-order
derivative expansion. At next-to-leading order, the flows of the wave function
renormalizations are described in terms of anomalous dimensions
\begin{equation}
	\eta_{\phi}=-\partial_t \mbox{ln} Z_{\phi,k},\ 
        \eta_{\psi}=-\partial_t \mbox{ln} Z_{\psi,k}\,.
\end{equation}
In order to fix the standard RG invariance of field rescalings, we define the
renormalized fields as
\begin{equation}
 	\tilde{\phi}=Z_{\phi,k}^{1/2}\phi,\ \tilde{\psi}=Z_{\psi,k}^{1/2}\psi.
\end{equation}
We introduce the dimensionless quantity
\begin{equation}
 	\tilde{\rho}=Z_{\phi,k}k^{-2}\rho\,,
\end{equation}
as well as the dimensionless renormalized Yukawa coupling and the
dimensionless potential,
\begin{equation}
 	h_k^2=Z_{\phi,k}^{-1}Z_{\psi,k}^{-2}\bar{h}^2_k,\quad
        u_k(\tilde\rho)= k^{-4}\,U_k(\rho)|_{\rho=k^2\tilde\rho/Z_{\phi,k}}\,.
\end{equation}
The full set of flow equations for this Yukawa-like ansatz and for an
arbitrary number of dimensions $d$ has been derived in \cite{Gies:2009}. For
our purposes, we specialize to $d=4$ and use a linear regulator function $R_k$
which is optimized for the present truncation \cite{Litim:2001up}. The flow of the
effective potential is given by
\begin{eqnarray}\label{basic:flowequation}
  \partial_t u_k(\tilde\rho) &=& -4 u_k + (2+\eta_{\phi})\tilde{\rho}u_k'\\
  & & +\frac{1}{16\pi^2}\left[\frac{(1-\frac{\eta_{\phi}}{6})}
    {2(1+u_k'+2\tilde{\rho}u_k'')}
    -\Nf\, \frac{2(1-\frac{\eta_{\psi}}{5})}{1+2\tilde{\rho}h_k^2}\right]\,. \nonumber
\end{eqnarray}
For the symmetric phase (SYM), we expand the effective potential around zero
field, 
\begin{eqnarray}\label{eq:symeffpot}
  u_k&=&\sum_{n=1}^{N_p}u_{n,k}\tilde{\rho}^n 
  = m_k^2\tilde{\rho}+\frac{\lambda_{2,k}}{2!}\tilde{\rho}^2
  +\frac{\lambda_{3,k}}{3!}\tilde{\rho}^3+...\nonumber. 
\end{eqnarray}

For the SSB phase, where the minimum of the effective potential $u_k$ acquires
a nonzero value $\kappa_k:=\tilde\rho_{\mbox{min}}> 0$, we use the expansion
\begin{eqnarray}
 	u_k&=&\sum_{n=2}^{N_p}u_{n,k}(\tilde{\rho}-\kappa_k)^n\\
	   &=&\!\frac{\lambda_{2,k}}{2!}(\tilde{\rho}-\kappa_k)^2
           \!+\frac{\lambda_{3,k}}{3!}(\tilde{\rho}-\kappa_k)^3+...%\nonumber\\
	  \label{eq:uexpSSB}.
\end{eqnarray}
Given the flow of $u_k$ \eqref{basic:flowequation}, the flows of $m_k^2$ or
$\lambda_{n,k}$ in both phases can be read off from an expansion of the flow
equation and a comparison of coefficients. For the flow of $\kappa_k$, we use
the fact that the first derivative of $u_k$ vanishes at the minimum,
$u_k'(\kappa_k)=0$. This implies
\begin{eqnarray}
 0= \pat u_k'(\kappa_k)&=&\partial_t u_k'(\tilde\rho)|_{\tilde\rho=\kappa_k}
  +(\partial_t \kappa_k)u_k''(\kappa_k)\nonumber\\
	\Rightarrow \partial_t \kappa_k&=&-\frac{1}{u_k''(\kappa_k)}\partial_t
        u_k'(\tilde\rho)|_{\tilde\rho=\kappa_k}\,. \label{eq:kappa}
\end{eqnarray}
The explicit flow equations for the running parameters will be given in the
following sections for the SYM and the SSB phase separately. For convenience,
we suppress the index $k$ from now on. 

Let us finally discuss several constraints on the couplings as, e.g., dictated
by our truncation. As our truncation is based on a derivative expansion,
satisfactory convergence is expected if the higher derivative operators take
little influence on the flow of the leading-order terms. In the present case,
the leading-order effective potential and Yukawa coupling receive higher-order
contributions only through the anomalous dimensions. Therefore, convergence of
the derivative expansion requires
\begin{equation}
 \eta_{\psi,\phi} \lesssim O(1).\label{eq:validity}
\end{equation}
This condition will serve as an important quality criterion for our
truncation. 

The SYM regime is characterized by a minimum of $u_k$ at vanishing field. A
simple consequence is that the mass term needs to be positive. Also, the
potential should be bounded from below, which in the polynomial expansion
translates into a positive highest nonvanishing coefficient,
\begin{equation}
m^2,\,  \lambda_{n_{\text{max}}}>0. \label{eq:constraint1}
\end{equation}
In the SSB regime, the minimum must be positive, $\kappa>0$, the potential
should be bounded, and in addition the potential at the minimum must have
positive curvature,
\begin{equation}
\kappa, \, \lambda_{n_{\text{max}}},\, \lambda_2>0.\label{eq:constraint2}
\end{equation}
Finally, Osterwalder-Schrader positivity requires 
\begin{equation}
h^2>0.\label{eq:hcrit}
\end{equation}
Beyond that, there are no constraints on the size of the couplings as in
perturbation theory. 

Let us finally mention that the derivative expansion is not only technically
motivated due to its systematics and consistency, it has also proved to
contain the relevant degrees of freedom for a variety of systems, yielding
quantitatively satisfactory results. RG flows of Yukawa systems have, for
instance, been successfully studied in QCD \cite{Jungnickel:1995fp}, critical phenomena
\cite{Rosa:2000ju}, and ultra-cold fermionic atom gases \cite{Birse:2004ha}.

\section{The Symmetric Regime (SYM)}
\label{sec:SYM}

In the following two sections, we carefully analyze the system in the SYM as
well as the SSB regime for $\Nf=1$. We have checked that our findings for the
search for non-Gau\ss ian fixed points extend also to larger natural values of
the flavor number. Fractional flavor numbers will be discussed separately,
below. 

In addition to the flow of the effective potential \eqref{basic:flowequation},
the flow equations for $h_k^2$ and the anomalous dimensions read in the SYM
regime \cite{Gies:2009}: 
\begin{eqnarray}\label{eq:symyuka}
 \partial_t h^2 &=&(\eta_\phi+2\eta_\psi)h^2\nonumber\\
	& &+\frac{h^4}{8\pi^2}\left(\frac{(1-\frac{\eta_\psi}{5})}{(1+m^2)}
          +\frac{(1-\frac{\eta_\phi}{6})}{(1+m^2)^2}\right),\label{eq:hNLO}\\
 \eta_\phi&=& \frac{h^2}{16\pi^2}(4-\eta_\psi),\label{eq:etaphiLO}\\
 \eta_\psi&=&
 \frac{h^2}{16\pi^2}\frac{(1-\frac{\eta_\phi}{5})}{(1+m^2)^2}.\label{eq:etapsiLO} 
\label{eq:anomaloussym}
\end{eqnarray}
For a different flavor number, the scalar anomalous dimension $\eta_\phi$
receives an additional factor of $\Nf$. 

\subsection{Fixed-point search to leading order}

For a first glance at the system, let us consider the leading-order derivative
expansion where the anomalous dimensions are set to zero, $ \eta_\phi,\
\eta_\psi =0$. For simplicity, let us furthermore neglect higher-order scalar
couplings, $\lambda_{\geq 3}=0$. Then we obtain a set of three coupled flow
equations for $m,\ \lambda_2$ and $h^2$,
\begin{eqnarray}
 \partial_t m^2 &=& -2m^2 + \frac{h^2}{4\pi^2}-\frac{3\lambda_2}{32\pi^2}\frac{1}{(1+m^2)^2}\,,\\
 \partial_t \lambda_2 &=& -\frac{h^4}{\pi^2}+\frac{9 \lambda_2^2}{16\pi^2}\frac{1}{(1+m^2)^3}\,,\\
 \partial_t h^2 &=& \frac{h^4}{8\pi^2}\frac{(2+m^2)}{(1+m^2)^2}\,. \label{eq:hLO}
\end{eqnarray}
Setting all flows to zero as required by a fixed point solution, we
immediately see from \Eqref{eq:hLO} that the only Yukawa fixed point is
$h^{\ast 2}=0$. This implies for the $\phi^4$ interaction that also
$\lambda_{2}^{\ast}=0$ and successively $m^\ast=0$. We conclude that the only
possible fixed point is the Gaussian fixed point. No traces of asymptotic
safety can be found in this case and the system is trivial if we insist on the
limit $\Lambda_{\text{UV,max}}\to \infty$.

A similar conclusion also applies to higher orders in the polynomial expansion
of the effective potential. As the flow equation for $h$ given in
\Eqref{eq:hLO} still remains valid, the Yukawa coupling remains trivial. No
asymptotically safe system with interacting fermions can be constructed in
this leading-order truncation.

\subsection{Fixed-point search at next-to-leading order}

At next-to-leading order, the anomalous dimensions need to be included. The
corresponding flow equations \eqref{eq:etaphiLO} and \eqref{eq:etapsiLO} are
purely algebraic and can be solved explicitly:
\begin{eqnarray}
 \eta_\phi&=&\frac{5 h^2 - 320 \pi^2 (1 + m^2)^2}{h^4 - 1280\pi^4(1 + m^2)^2
 }h^2, \label{eq:etaphiSYM}\\
 \eta_\psi&=&\frac{4 h^2 - 80 \pi^2}{h^4 - 1280\pi^4 (1 + m^2)^2
 }h^2\,\label{eq:etapsiSYM}. 
\end{eqnarray}
Plugging these solutions into the flow of the Yukawa coupling \eqref{eq:hLO},
the fixed point equation $\pat h^2=0$ can be solved algebraically, indeed
revealing a possible fixed point,
\begin{eqnarray}\label{eq:symyukafix}
 h^{*2} &=& \frac{40\pi^2}{11 + 6 m^2}\bigg[(-65 - 124 m^2 - 59 m^4)+(1 + m^2)\nonumber \\
	&\times& \sqrt{5545 + 9710 m^2 + 4729 m^4 + 288 m^6}\bigg]\,,
\end{eqnarray}
where we have dropped the solutions with $h^2 \leq 0$. Of course,
Eq. \eqref{eq:symyukafix} describes a true fixed point, only if $m$ also
approaches a fixed point. In the following, we will call these conditional
fixed-point solutions \emph{pseudo-fixed points}. Next, we analyze the flow
equation for the mass $m$ which is obtained from \Eqref{basic:flowequation}
without any further approximation:
\begin{equation}
 \partial_t m^2=(-2+\eta_\phi)m^2+\frac{h^2}{4\pi^2}(1-\frac{\eta_\psi}{5})
-\frac{3\lambda_2}{32\pi^2}\frac{1-\frac{\eta_\phi}{6}}{(1+m^2)^2}\,.
\end{equation}
Solving the fixed-point condition $\pat m^2=0$ yields a pseudo-fixed point value as a
function of the $\phi^4$ interaction, $m^{\ast 2}=m^{\ast 2}(\lambda_2)$. This
hierarchy of fixed-point conditions persists to all orders of the expansion of
the effective action: solving the fixed-point condition from the flow of
$\lambda_2$, $\pat \lambda_2=0$, yields a pseudo-fixed point value as a function of
the next coupling, $\lambda_2^\ast=\lambda_2^\ast(\lambda_3)$, or more
generally, $\lambda_n^\ast=\lambda_n^\ast(\lambda_{n+1})$. This seems to
suggest a method to construct a fixed point potential for this Yukawa system. 

However, let us take a closer look at the simplest approximation where we
neglect all higher couplings $\lambda_{\geq 3}=0$. In this case, the hierarchy
truncates and all fixed-point conditions can be resolved.  A numerical
solution of the resulting fixed-point equations reveals two solutions, true fixed points which
are listed in Tab. \ref{tab:anomdimsym0}

\begin{table}[ht]
\footnotesize
\centering
	\begin{tabular}{c|c|c|c|c|c}
	FP&$m^\ast$&	$h^{\ast 2}$	&$\lambda_2^\ast$	  & $\eta_{\psi}^\ast$ & $\eta_{\phi}^\ast$ \\
	\cline{1-6}
	1&41.86	&188906	&$-8.4111*10^6$	& 4.0216 &-25.8847\\
	2&403.84	&$7.6573*10^6$	&$1.9079*10^{10}$	& 4.0013 &-62.6194\\
	\end{tabular}
\caption{Fixed-point values for the set of equations for $h^2,\
          \lambda_2$, fully including the anomalous dimensions. The
          $\eta_{\phi,\psi}^\ast$ values are unacceptably large, indicating that
          these solutions are outside the validity range of the derivative
          expansion. }
\label{tab:anomdimsym0}
\end{table}

The numerical values for the anomalous dimensions are clearly outside the
validity regime of the derivative expansion, cf. \Eqref{eq:validity}. Also for
the first fixed point, the resulting effective potential is unstable. We
conclude that none of the fixed points satisfies our acceptance criteria,
i.e. they are either physically irrelevant or could possibly be artifacts of
the truncation.

In order to explore the effect of including a full effective potential, we
could perform the same analysis for higher and higher orders in the Taylor
expansion of the effective potential. Here, we propose a simpler method for
analyzing the system: for this, we observe that the above set of equations is
actually exact except for the unknown coupling $\lambda_3$ that was set to
zero above. Instead of trying to find an approximate value for $\lambda_3$, we
simply allow for any value in order to obtain a non-Gau\ss ian fixed point in
the flows of $m^2,h^2,\lambda_2$ and in the anomalous dimensions. Indeed, we
numerically find a line of pseudo-fixed points parameterized by $\lambda_3$ (if these
fixed points were real, the correct $\lambda_3$ value would be singled out by
its flow equation). In table \ref{tab:anomdimsym}, we give a list of these pseudo-fixed 
points for squared mass values between 0 and 100.

The conclusion drawn from this list is similar to the one above found in the
$\phi^4$ truncation: the anomalous dimensions for all fixed points are too
large for being acceptable. For larger mass values, the required $\lambda_3$
value even turns negative (even though this could be compensated by higher
couplings).  We conclude that even if the true value for $\lambda_3$ would
lie in the required range for a fixed point, this fixed point is likely to be
an artifact of the truncation. 

In summary, we do not find any reliable evidence for a non-Gau\ss ian fixed
point in the SYM regime of this simple Yukawa system. The $\Nf=1$ system in
the SYM regime shows no sign of asymptotic safety. We have also found no
indications for asymptotic safely for larger flavor number in the SYM regime. 

\begin{table}[ht]
\footnotesize
\centering
	\begin{tabular}{c|c|c|c|c|c}
	$m^{\ast 2}$&	$h^{\ast 2}$	&$\lambda_2^\ast$	& $\lambda_3$ & $\eta_{\psi}^\ast$ & $\eta_{\phi}^\ast$\\
	\cline{1-6}
	0	&339	&-592		&$2.3*10^{6}$	&-20.7	& 53.3\\
	0.01	&343	&-607		&$2.3*10^{6}$	&-21.8	& 56.0\\
	0.1	&378	&-747		&$3.2*10^{6}$	&-36.9	& 98.0\\
	0.5	&562	&-1663		&$1.1*10^{7}$	&23.2 	& -68.3\\	
	1	&854	&-3653		&$3.6*10^{7}$	&9.7 	& -31.0\\
	5	&5570	&-81688		&$5.5*10^{9}$	&4.5	& -18.0\\
	10	&16776	&-425265	&$8.0*10^{10}$	&4.2	& -18.8\\
	100	&807421	&$+5.7*10^{7}$	&$-1.5*10^{15}$	&4.0	& -35.0\\
	\end{tabular}
        \caption{Fixed-point values for the set of equations for $m^2,\ h^2,\
          \lambda_2$ for a given value of $\lambda_3$ , fully including the anomalous dimensions. The
          $\eta_{\phi,\psi}$ values are again unacceptably large, indicating that
          these solutions are outside the validity range of the derivative
          expansion.} 
\label{tab:anomdimsym}
\end{table}

\section{The Regime of Spontaneous Symmetry Breaking (SSB)}
\label{sec:SSB}

The fact that no acceptable fixed point was found in the SYM regime can be
traced back to the properties of the Yukawa coupling flow. At leading-order
derivative expansion, cf. \Eqref{eq:hLO}, the Yukawa flow only supports the
Gau\ss ian fixed point, implying a trivial interaction. In order to circumvent
this no-go property, negative terms have to occur on the right-hand side of
this equation. In the SYM regime, this can only be induced by the anomalous
dimension terms in \Eqref{eq:hNLO} which have to be sizeable and thus
unacceptably large.

A new possibility opens up in the SSB regime. Here, further effective
interactions arise owing to possible couplings to the condensate. On the other
hand, the flow in the SSB regime generically has the tendency to induce
decoupling of massive modes and thus an unwanted freeze-out of the flow. This
problem is automatically avoided if the expectation value of the field
exhibits a conformal behavior. Therefore, we analyze the SSB flow in the
following with an emphasis on the running of the expectation value.

From the effective potential \eqref{basic:flowequation}, together with
\Eqref{eq:kappa}, the flow of the dimensionless expectation value $\kappa$ and
the scalar couplings occurring in the expansion \eqref{eq:uexpSSB} can be
derived:
\begin{eqnarray}
  \partial_t
  \kappa&=&-(2+\eta_{\phi})\kappa-\frac{h^2}{4\pi^2}
\frac{(1-\frac{\eta_{\psi}}{5})}{\lambda_2(1+2 h^2\kappa)^2}\nonumber\\ 
&+& \frac{(3\lambda_2+2\kappa\lambda_3)}{32\pi^2}
\frac{(1-\frac{\eta_{\phi}}{6})}{\lambda_2(1+2\kappa\lambda_2)^2}\,,\label{eq:kappaSSB}\\
\partial_t\lambda_2&=& \lambda_3\partial_t\kappa + 2\eta_{\phi}\lambda_2+
(2+\eta_{\phi})\kappa\lambda_3 \label{eq:lam2SSB}\\ 
&& -\frac{h^4}{\pi^2}\frac{(1-\frac{\eta_{\psi}}{5})}{(1+2\kappa
  h^2)^3}
+ \frac{(3\lambda_2+2\kappa\lambda_3)^2}{16\pi^2}
\frac{(1-\frac{\eta_{\phi}}{6})}{(1+2\kappa\lambda_2)^3} \nonumber\\
&&-\frac{5\lambda_3}{32\pi^2}\frac{(1-\frac{\eta_{\phi}}{6})}{(1+2\kappa\lambda_2)^2},
\nonumber\\
\partial_t \lambda_3 &=& (2+3\eta_{\phi})\lambda_3
-\frac{3(3\lambda_2+2\kappa\lambda_3)^3}{16\pi^2}
\frac{(1-\frac{\eta_{\phi}}{6})}{(1+2\kappa\lambda_2)^4}\label{eq:l3LOSSB}\\
&+&\frac{6h^6}{\pi^2}\frac{(1-\frac{\eta_{\psi}}{5})}{(1+2\kappa h^2)^4}
+\frac{15\lambda_3(3\lambda_2+2\kappa\lambda_3)
(1-\frac{\eta_{\phi}}{6})}{16\pi^2(1+2\kappa\lambda_2)^3}\nonumber.  
\end{eqnarray}
In the last equation, we have omitted the contributions from $\lambda_4$ for
simplicity. The flow of the Yukawa coupling is \cite{Gies:2009}:
\begin{eqnarray}\label{eq:ssbyuk}
 \partial_t h^2 &=& (\eta_{\phi}+2\eta_{\psi})h^2\nonumber\\
&+& \frac{h^4}{8\pi^2}\frac{\left(\frac{1-\frac{\eta_{\psi}}{5}}{1+2\kappa
      h^2}
+\frac{1-\frac{\eta_{\phi}}{6}}{1+2\kappa \lambda_2} \right)}
{(1+2\kappa h^2)(1+2\kappa\lambda_2)} \nonumber\\
&-& \frac{h^4(3\kappa\lambda_2+2\kappa^2\lambda_3)}{4\pi^2}
\frac{\left(\frac{1-\frac{\eta_{\psi}}{5}}{1+2\kappa h^2}
+\frac{2(1-\frac{\eta_{\phi}}{6})}{1+2\kappa \lambda_2} \right)}
{(1+2\kappa h^2)(1+2\kappa\lambda_2)^2} \nonumber\\
&-&
\frac{h^6\kappa}{2\pi^2}\frac{\left(\frac{2(1-\frac{\eta_{\psi}}{5})}{1+2\kappa
      h^2}
+\frac{1-\frac{\eta_{\phi}}{6}}{1+2\kappa \lambda_2} \right) }{(1+2\kappa
h^2)^2(1+2\kappa\lambda_2)}\label{eq:hqSSB}. 
\end{eqnarray}
In comparison with the SYM regime \eqref{eq:hNLO}, we observe a significantly
more complex structure here which is induced by the fact that the fluctuations
propagate in a nonvanishing field expectation value. This is also true for the
flow of the wave function renormalizations in terms of the anomalous
dimensions
\begin{eqnarray}
 \eta_{\phi}&=& \frac{h^2}{8\pi^2}\frac{(1-\eta_{\psi})}{(1+2\kappa h^2)^3}
-\frac{h^2}{16\pi^2}\frac{(2-\eta_{\psi})}{(1+2\kappa h^2)^2}\label{eq:anomalous}\\
&+&
\frac{\kappa(3\lambda_2+2\kappa\lambda_3)^2}{16\pi^2}\frac{1}{(1+2\kappa\lambda_2)^4}
+\frac{h^2}{4\pi^2}\frac{(1-2\kappa h^2)}{(1+2\kappa h^2)^4}\,,\nonumber\\
 \eta_{\psi}&=& \frac{h^2}{16\pi^2}\frac{(1-\frac{\eta_{\phi}}{5})}
{(1+2\kappa h^2)(1+2\kappa\lambda_2)^2},\label{eq:anomalouspsi}
\end{eqnarray}
which should be compared to Eqs. \eqref{eq:etaphiLO} and
\eqref{eq:etapsiLO} in the SYM regime. For other flavor numbers, all terms
proportional to $h^2$ in the $\eta_\phi$ equation receive an additional factor
of $\Nf$. 

\subsection{Fixed-point search to leading order}\label{basic}

Let us again start with the leading-order derivative expansion, $ \eta_\phi,\
\eta_\psi =0$, confining ourselves initially to the $\phi^4$ truncation with
three parameters $\kappa,h^2,\lambda_2$ and all other $\lambda_{\geq 3}=0$. 

We first consider the two fixed-point equations $\pat h^2=0$ and $\pat
\lambda_2=0$, and let the value for $\kappa$ be undetermined for the
moment. The resulting equations can be solved analytically, revealing a set of
solutions.  Only one of them fulfills the requirement $\lambda_2>0$ and $h^2
>0$. This solution reads
\begin{eqnarray}
  h^{ \ast 2} &=& \frac{0.0616116}{\kappa}, \quad
  \lambda_2^{\ast} = \frac{0.0880079}{\kappa}\,.
\end{eqnarray}
Plugging these pseudo-fixed-point values into the fixed-point equation for
$\kappa$, we find
\begin{equation}
 \partial_t \kappa = -2\kappa - 0.0071873 \neq 0 \ \mbox{for } \kappa
 >0\,.\label{eq:kapFP} 
\end{equation}
This proves that there is no nontrivial fixed point at leading-order
derivative expansion in the $\phi^4$ truncation for $\Nf=1$. This conclusion
holds also for all higher flavor numbers.

In order to test the reliability of the $\phi^4$ truncation, we next include
the $\lambda_3$ coupling and its flow equation \eqref{eq:l3LOSSB} still
keeping the anomalous dimensions at zero. Unfortunately, we have not been able
to find an analytical solution of the resulting fixed point equations. Hence,
we start numerically with the same strategy as given above: for a given
(arbitrary) value of $\kappa$, we solve the fixed-point equations for
$\lambda_2,\lambda_3$ and $h^2$. For the numerics, we choose values of
$\kappa$ ranging from 10 down to 0.001.

\begin{table}[ht]
\footnotesize
\centering
	\begin{tabular}{c|c|c|c|c|c}
	$\kappa$&	$h^{\ast 2}$	&$\lambda_2^{\ast}$	&$\lambda_3^{\ast}$	& $\partial_t\kappa$ & $\partial_t\kappa_{\phi^4}$\\
	\cline{1-6}
	10	&0.00616	&0.00880	&$4.67*10^{-8}$	&-20.0  & -20.0\\
	1	&0.0615		&0.0880		&$4.67*10^{-5}$	&-2.01  & -2.01\\
	0.1	&0.609		&0.882		&$4.61*10^{-2}$	&-0.210 & -0.207\\	
	0.01	&5.57		&8.96		&38.5		&-0.026 & -0.027\\
	0.001	&47.1		&91.0		&10800		&-0.0056& -0.0092\\
	0.0001	&451		&912		&$1.26*10^{6}$	&-0.0033& -0.0074\\
        0.00001	&4490		&9120		&$1.29*10^{8}$	&-0.0031& -0.0072\\
	\end{tabular}
        \caption{Pseudo-fixed-point values for the fixed-point equations for
          $h^2,\ \lambda_2$ and $\lambda_3$. These values would correspond to
          a true non-Gau\ss ian fixed point, if $\partial_t \kappa$ eventually
          went to zero, which is not the case here.  The index $\phi^4$ in the
          last row denotes the values obtained in a pure $\phi^4$ truncation.} 
\label{tab:extendedtrunc}
\end{table}
 
% \begin{figure}[ht]
%  \centering
% 	\includegraphics[width=0.40\textwidth]{couplings.eps}
% 	\caption{Comparison of the values for $\partial_t \kappa$ in the two truncations. The red line gives the extended truncation including $\lambda_3$, the dashed black line is the basic truncation.}
% 	\label{fig:kappa}
% \end{figure}
We find a large number of pseudo-fixed-point solutions
$\left\lbrace h^{\ast 2}, \lambda_2^{\ast}, \lambda_3^{\ast}\right\rbrace $
for this system ($\sim 40$ depending on the value of $\kappa$), but again only
one of these solutions given in table \ref{tab:extendedtrunc} fulfills the
necessary criteria for the parameters,
cf. Eqs.~\eqref{eq:validity}-\eqref{eq:hcrit}.

As is visible in table \ref{tab:extendedtrunc}, the values of the $h^{\ast 2}$ and
$\lambda_2^{\ast}$ are only slightly changed compared to the $\phi^4$ truncation for
$\kappa$ between $10$ and $0.01$. We still observe a scaling $h^{\ast 2},\lambda_2^{\ast}
\propto 1/\kappa$ in this regime, so that the effect of $\lambda_3$ can be
considered as a small perturbation. The value for $\lambda_3^{\ast}$ approximately
scales as $1/\kappa^3$. For even smaller values of $\kappa$ ($\kappa < 0.01$),
$\lambda_3^{\ast}$ grows large and exerts a quantitative influence on the
flow. Nevertheless, we still find no indications for a non-Gau\ss ian fixed
point in the $\kappa$ flow. Even for the smallest $\kappa$ values which are
accessible by our numerics, $\kappa\sim 0.00001$, we observe a monotonous
increase of $\pat\kappa/\kappa$, whereas an approach to a fixed point would
require a decrease of this combination towards zero. 

We conclude that we find no evidence for asymptotic safety of our Yukawa model
at leading order in the derivative expansion in the SSB regime for
$\Nf=1$. This conclusion extends also to larger values of $\Nf$.

\subsection{Fixed-point search including anomalous dimensions}

The inclusion of the anomalous dimensions is, in principle, straightforward. 
The corresponding equations for $\eta_\psi$ and $\eta_\phi$
\eqref{eq:anomalous} and \eqref{eq:anomalouspsi} can be solved analytically;
their solution can then be plugged into the flows of $\kappa$, $h^2$,
$\lambda_2$, \dots which can be treated as before. 

In practice, the inclusion of the anomalous dimensions increases the
nonlinearities of the flow equations substantially. This holds already for the
$\phi^4$ truncation. For our simplified strategy of first fixing a value of
$\kappa$ and then solving for possible pseudo-fixed point values of $h^2$ and
$\lambda_2$, an exhaustive search of all possible pseudo-fixed point values
appears numerically not feasible. Hence, we have used an algorithm that
searches for one pseudo-fixed point of $h^2$ and $\lambda_2$ for a given value
of $\kappa$ and given seed values $h^2_{\text{seed}}$ and
$\lambda_{2,\text{seed}}$. We have chosen the seed values on a grid of
reasonable values compatible with the constraints
\eqref{eq:constraint1}-\eqref{eq:hcrit}.  This procedure leads to the results
shown in table \ref{tab:anomdim}.

\begin{table}[ht]
\footnotesize
\centering
	\begin{tabular}{c|c|c|c|c|c}
	$\kappa$&	$h^{\ast 2}$	&$\lambda_2^{\ast}$	&$\eta_{\phi}^{\ast}$	& $\eta_{\psi}^{\ast}$ & $\partial_t\kappa_t$\\
	\cline{1-6}
	100	&0.00185	&0.00210	&$1.11*10^{-5}$	&$4.24*10^{-6}$	& -200.008\\
	10	&0.0185		&0.0210		&$1.11*10^{-4}$	&$4.24*10^{-5}$	& -20.008\\
	1	&0.185		&0.210		&$1.11*10^{-3}$	&$4.24*10^{-4}$	& -2.008\\
	0.1	&1.85		&2.10		&0.0111		&$4.23*10^{-3}$ & -0.2083\\	
	0.01	&18.5		&21.1		&0.111		&0.0413 	& -0.0282\\
	0.001	&181.7		&221.0		&1.07		&0.32		& -0.0098\\
	0.0001	&1516		&1602		&1.73		&2.98		& -0.0027\\
	0.00001	&12030		&9778		&4.42		&3.90		& -0.0016\\
	0.000001&68411		&93113		&4.92		&4.16		& -0.0012\\
	\end{tabular}
        \caption{Pseudo-fixed-point values for the fixed-point equations for
          $h^2,\ \lambda_2$ including anomalous dimensions. As in table
          \ref{tab:extendedtrunc}, these values would correspond to
          a true non-Gau\ss ian fixed point, if $\partial_t \kappa$ eventually
          went to zero, which is not the case here.}  
\label{tab:anomdim}
\end{table}

In the regime of small anomalous dimensions ($\eta_{\phi,\psi} < 1$), we again
observe a $1/\kappa$-behavior for the pseudo-fixed-point values of $\lambda$
and $h^2$,
\begin{eqnarray}
 h^{\ast 2}=\frac{0.185059}{\kappa}\\
 \lambda^{\ast}=\frac{0.21018}{\kappa}\,.
\end{eqnarray}
An inclusion of the anomalous dimensions does not lead to a vanishing of $\partial_t
\kappa$ within the range of $\kappa$ values studied here. Also, the ratio
$\pat\kappa/\kappa$ does not exhibit a tendency towards zero as it should if a
fixed point existed in the vicinity of the present parameter range.

In contrast to the numerical procedures used above, the algorithm used for the
study in this section does not exhaust all possible pseudo-fixed-point
values. Hence, it is conceivable that a fixed point with suitable properties
still exists but has evaded detection due to numerical reasons. Our studies,
however, do not give any indication for such a scenario. 

\section{Flow equation for Halpern-Huang potentials}
\label{sec:HH}

A special case of the asymptotic-safety scenario is given by an asymptotically
free theory where the Gau\ss ian fixed point has at least one relevant (or
marginally relevant) direction inducing an interacting theory. In the case of
a pure scalar system, this would correspond to an asymptotically free analogue
of $\phi^4$ theory. In a local-potential approximation (leading-order
derivative expansion), a linearized analysis in the Gau\ss ian fixed-point
regime reveals that relevant directions exist in the form of nonpolynomial
potentials \cite{Halpern:1995vf}. For these Halpern-Huang potentials, the Gau\ss ian fixed
point is asymptotically free, hence these potentials have been proposed as a
candidate for a nontrivial Higgs sector \cite{Halpern:1995vf}. A large-$N$
solution of the flow towards the strongly interacting low-energy sector of
these theories revealed that also the hierarchy problem can be softened
considerably \cite{Gies:2000xr}.\footnote{A drawback of these relevant
  directions is that there exist infinitely many, implying an infinity of
  physical parameters \cite{Morris:1995af}. Nevertheless, the theory can be
  predictive in the sense that the initial conditions are not given by
  finitely many parameters but by a full function. }

Here, we want to study whether the Halpern-Huang directions can be used as an
ingredient of an asymptotically safe Yukawa system in combination with a
fermionic sector. For this, the Halpern-Huang solutions need to persist also
for finite Yukawa coupling and should go along with a non-Gau\ss ian fixed
point of the Yukawa coupling.

The linearized flow equation for the effective potential in the vicinity of
the Gau\ss ian fixed point (Eq. (\ref{basic:flowequation})) reads in $d=4$ and
for $\Nf=1$:
\begin{eqnarray}\label{eq:linflow}
 \partial_t u_k &=& -4 u_k + (2+\eta_\phi)\tilde\rho u'_k -
 \frac{(1-\frac{\eta_\phi}{6})}{32\pi^2}(u'_k+2\tilde\rho u''_k)\nonumber\\ 
		& &
                +\frac{(1-\frac{\eta_\psi}{5})}{4\pi^2}\frac{\tilde\rho h^2}{1+2\tilde\rho
                  h^2}\,. 
\end{eqnarray}
Here, we have shifted the irrelevant field-independent part of the potential
by a constant in order to remove the constant contributions in the fermionic
term. For $h^2=0$, we rediscover the Halpern-Huang equation. Now, assuming
that $h^2$ approaches a non-Gau\ss ian fixed point $h^{\ast 2}$ in the UV, we
observe that the Gau\ss ian fixed point for the potential $u_k=0$ is no longer
a solution due to the presence of the fermionic contribution.  Therefore, the
fermionic inhomogeneity implies that the effective potential $u_k$ can no
longer be asymptotically free. If on the other hand $h^2 =0$ in the UV, the
Yukawa coupling stays zero during the flow, as is visible from
\Eqref{eq:ssbyuk}. We conclude that there is no nontrivial generalization of
the Halpern-Huang idea to the Yukawa system in the truncation considered here.  

\section{Yukawa systems at small $\Nf$}
\label{sec:Nf}

Since no acceptable asymptotically safe Yukawa system has been found in the
studies above, the question arises as to whether there is a fundamental obstacle
against a conformal behavior of the Higgs vacuum expectation value, or whether
the class of simple systems has been too restrictive so far. In the following,
we demonstrate that the latter is the case by explicitly demonstrating the
existence of an asymptotically safe Yukawa system at small $\Nf$. 

A small-$\Nf$ study can be motivated from the fact that the conformal
behavior of the scalar expectation value is inhibited by the dominance of the
fermionic fluctuations in the $\kappa$ flow, as is visible, e.g., in
Eqs. \eqref{eq:kappaSSB} and \eqref{eq:kapFP}. This fermionic dominance is the
origin of the negative sign in front of the constant term in \Eqref{eq:kapFP}
which shifts a possible fixed-point value to physically unacceptable negative
values of $\kappa$. Since fermionic loops are proportional to the number of
flavors, this fermionic dominance can be expected to vanish for small $\Nf$. 

At arbitrary flavor number, the relevant flow equations in the simple $\phi^4$
truncation in the SSB regime read to leading-order in the derivative expansion
\begin{eqnarray}
 \partial_t \kappa&=&-2\kappa-\frac{\Nf h^2}{4\pi^2}\frac{1}{\lambda_2(1+2
   \kappa h^2)^2}\label{basickappa}\\ 
 & & +\frac{3}{32\pi^2}\frac{1}{(1+2\kappa\lambda_2)^2}\,,\nonumber\\
\partial_t\lambda_2&=&-\frac{\Nf h^4}{\pi^2}\frac{1}{(1+2\kappa
  h^2)^3}+\frac{9\lambda_2^2}{16\pi^2}\frac{1}{(1+2\kappa \lambda_2)^3}\,. 
\end{eqnarray}
The flow of $h^2$ remains unchanged as no closed fermion loop contributes to
it. Following our strategy introduced above, we first pick a value for
$\kappa$ and then solve the fixed-point equations of $h^2$ and $\lambda_2$,
yielding pseudo-fixed-point values $h^{\ast 2}$ and $\lambda_{2}^\ast$. Plugging
these values into the $\kappa$ flow, results in 
\begin{equation}
 \partial_t \kappa = -2\kappa + c(\Nf)\,,
\end{equation}
where $c(\Nf)$ denotes an $\Nf$-dependent constant. As observed above,
$c(\Nf)$ is negative for $\Nf=1$ or larger, cf. \Eqref{eq:kapFP}, which
corresponds to the observed absence of an acceptable fixed point. Decreasing
$\Nf$, $c(\Nf)$ becomes positive for $\Nf \lesssim 0.25$ and so gives rise to a
true non-Gau\ss ian fixed point $\kappa^\ast=c(\Nf)/2$ within this $\phi^4$
truncation. This demonstrates by way of example that a conformal behavior of
the Higgs expectation value is indeed possible as a matter of principle and
can support asymptotic safety of Yukawa systems. In the following, we study
the properties and consequences of this non-Gau\ss ian fixed point.

\subsection{Fixed-point analysis at leading order}

In order to illustrate the fixed-point properties, we choose $\Nf=1/10$ for
the sake of definiteness. The non-universal fixed-point values read
\begin{eqnarray}
 && \kappa^*=0.00165, \quad 
 \lambda_2^*=27.26,\quad
 h^{\ast 2}=81.13.
\end{eqnarray}
The universal critical exponents can be deduced from the linearized flow
around this fixed point, 
\begin{eqnarray}
 \partial_t \kappa &=&
 \mbox{B}_{\kappa}{}^{\kappa}(\kappa-\kappa^*)+\mbox{B}_{\kappa}{}^{\lambda}(\lambda_2-\lambda_2^*)+\mbox{B}_{\kappa}{}^{
   h}(h^2-h^{\ast 2})\nonumber\\
\partial_t \lambda_2 &=&
\mbox{B}_{\lambda}{}^{\kappa}(\kappa-\kappa^*)+\mbox{B}_{\lambda}{}^{\lambda}(\lambda_2-\lambda_2^*)+\mbox{B}_{\lambda}{}^{
  h}(h^2-h^{\ast 2})\nonumber\\
\partial_t h^2 &=& \mbox{B}_{h}{}^{\kappa}(\kappa-\kappa^*)+\mbox{B}_{h}{}^{
  \lambda}(\lambda_2-\lambda_2^*)+\mbox{B}_{h }{}^{h}(h^2-h^{\ast 2})\,,\nonumber
\end{eqnarray}
where we have dropped terms of higher order in the fixed-point distance. The
expansion coefficients  $\mbox{B}_i{}^j$ form the stability matrix $B$, the
negative eigenvalues $-\Theta_i$ of which correspond to the critical exponents. 
For $\Nf=1/10$, we find
\begin{eqnarray}
 &&\Theta_1=2.372,\quad
 \Theta_2=0.592, \quad
 \Theta_3=-2.859.
\end{eqnarray}
We observe two relevant directions, corresponding to $\Theta_{1,2}>0$, i.e.,
the universality class of the theory defined by this fixed point has two
physical parameters.

As a first check of this simple approximation, we determine the size of the
anomalous dimensions by inserting the fixed-point values into the
$\Nf$-generalized equations for $\eta_\psi$ and $\eta_\phi$. Neglecting the
back-reaction onto the flow of the couplings due to RG improvement, we indeed
obtain satisfactorily small values
\begin{eqnarray}
 &&\eta_\phi = 0.09,\quad
 \eta_\psi = 0.03,
\end{eqnarray}
which can be taken as an indication for the self-consistency of the derivative
expansion.

\subsection{Fixed-point analysis at next-to-leading order}

A full treatment of the anomalous dimensions indeed leads to the same
qualitative behavior. The critical fermion number slightly increases: below
$\Nf\lesssim 0.33$, a non-Gau\ss ian fixed point exists, the coupling values
of which are shifted as compared to the leading-order approximation. For
$\Nf=1/10$, we find
\begin{eqnarray}
 &&\kappa^\ast=0.00163, \quad
 \lambda_2^\ast=42.77,\quad
  h^{\ast 2}=191.22\,,\nonumber\\
 &&\eta_\phi^\ast = 0.086, \quad \eta_\psi^\ast = 0.565.
\end{eqnarray}
The coupling fixed points $\lambda_2^\ast$ and $h^{\ast 2}$ as well as the
fermion anomalous dimension exhibit larger variations. Reliable quantitative
predictions hence may require the inclusion of further boson-fermion
interaction terms in the truncation.  By contrast, the predictions for the
$\kappa$ fixed point and the scalar anomalous dimension are little
affected. We conclude that our result of a conformal behavior of the Higgs
expectation value is robust in the derivative expansion.

The number of relevant directions at the fixed point is also robust: two
critical exponents have a positive real part. In comparison with the
leading-order result, the $\Theta_i$ values not only vary quantitatively, but
also the relevant directions are now described by a pair of complex $\Theta_i$,
\begin{eqnarray}
 &&\Theta_{1,2}=1.619\mp 0.280 i,\quad
 \Theta_3=-3.680,
\end{eqnarray}
implying that the UV flow spirals out of the fixed-point region towards the
IR. 

\section{Predictive power of asymptotically-safe Yukawa systems}
\label{sec:PP}

\subsection{UV fixed-point regime and mass hierarchy}

The present simple Yukawa system for a small number of fermion flavors $\Nf
\lesssim 0.3$ supports a non-Gau\ss ian fixed point, rendering the system
asymptotically safe. The predictive power of these systems can be read off
from the critical exponents $\Theta_i$, as computed above for the special case
of $\Nf=1/10$. The number of positive $\Theta_i$ corresponds to the number of
physical parameters. The value of the real part of the largest $\Theta_i$ is
a measure for the severeness of the hierarchy problem;
e.g. $\Theta_{\text{max}}=2$ in the standard model at the Gau\ss ian fixed
point.  

\begin{figure}[ht]
\centering
\includegraphics[width=0.45\textwidth]{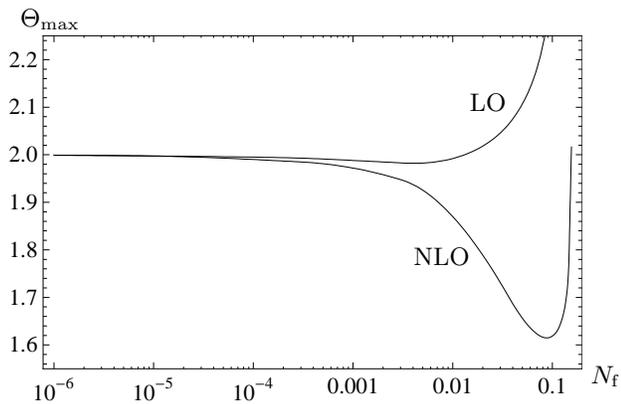}
\caption{$\Theta_{\text{max}}$ vs. $\Nf$ as a measure for the hierarchy
  problem.  For $\Nf\to 0$, the critical exponent approaches the
  standard-model value $\Theta_{\text{max}}=2$. The upper/lower curve
  corresponds to leading-order/next-to-leading-order derivative expansion.}
\label{fig:theta}
\end{figure}

Repeating the above analysis of the eigenvalues of the stability matrix $B$
for various values of $\Nf$ yields the values of $\Theta_{\text{max}}$ as
displayed in Fig.~\ref{fig:theta}. In the limit $\Nf\to 0$, the standard-model
value $\Theta_{\text{max}}=2$ is approached (even though the system is not in
the standard-model universality class, but at the non-Gau\ss ian fixed
point). For increasing $\Nf$, $\Theta_{\text{max}}$ decreases and approaches a
minimum. At next-to-leading order, the minimum is indeed close to $\Nf=1/10$
where $\Theta_{\text{max}}\simeq 1.6$. Here, the hierarchy problem is not
really overcome but at least somewhat reduced compared to the standard
model. Beyond this minimum, $\Theta_{\text{max}}$ increases again and
eventually diverges at the critical flavor number.

\subsection{Higgs mass from asymptotic safety}

For all values of $\Nf$ below the critical value, we find that two of the
critical exponents are positive (relevant eigendirection), implying that the
number of physical parameters of these systems is 2. As the SSB phase is
characterized by 3 standard-model parameters, top mass $m_{\text{top}}$, Higgs
mass $m_{\text{Higgs}}$ and vacuum expectation value $v$, only two of them are
needed to fix the Yukawa system at the non-Gau\ss ian fixed point. The third
parameter, say the Higgs mass, is a true prediction of the asymptotically safe
model. 

The physical parameters are related to the renormalized couplings $\kappa,h^2,
\lambda_2$ by
\begin{equation}\label{eq:topmass}
v=\lim_{k\to0} \sqrt{2\kappa}k, \quad m_{\text{top}}=\sqrt{ h^2 } v
, \quad 
 m_{\text{Higgs}}=\sqrt{ \lambda_2 } v.
\end{equation}
An asymptotically safe RG trajectory has to emanate from the non-Gau\ss ian
fixed point in the deep UV. The possible trajectories are further constrained
by the necessity that the system ends up in the SSB phase in order to describe
chiral-symmetry breaking and mass generation. 

In practice, we start the flow at a UV scale $\Lambda_{_\text{UV}}$ in the
vicinity of the non-Gau\ss ian fixed point, say at
$(\kappa^{\ast}+\delta\kappa,\lambda_2^{\ast},h^{2\ast}+\delta
h^2)$.\footnote{For generic perturbations $\delta\kappa$ and $\delta h^2$,
  this starting point is not in the critical surface spanned by the relevant
  directions. However, as the irrelevant perturbations around the fixed point
  die out rapidly towards the IR, the IR physics is purely determined by the
  projection of these perturbations onto the critical surface.} Then, the
perturbations $\delta\kappa$ and $\delta h^2$ are tuned such that the top mass
and the vacuum expectation value match their desired physical IR values. 
The resulting Higgs mass then is a parameter-free prediction of the system. 

This can be explicitly checked by varying the starting point, i.e., adding
irrelevant directions, under the constraint of keeping $v$ and
$m_{\text{top}}$ fixed. As another check of asymptotic safety, the UV scale
$\Lambda_{_\text{UV}}$ can be changed. The constraint of keeping $v$ and
$m_{\text{top}}$ fixed corresponds to staying on the line of constant
physics. Asymptotic safety then guarantees that the Higgs mass remains the
same and that the starting point approaches the non-Gau\ss ian fixed point for
increasing $\Lambda_{\text{UV}}$.

\begin{figure}
\begin{center}
 \includegraphics[width=0.35\textwidth]{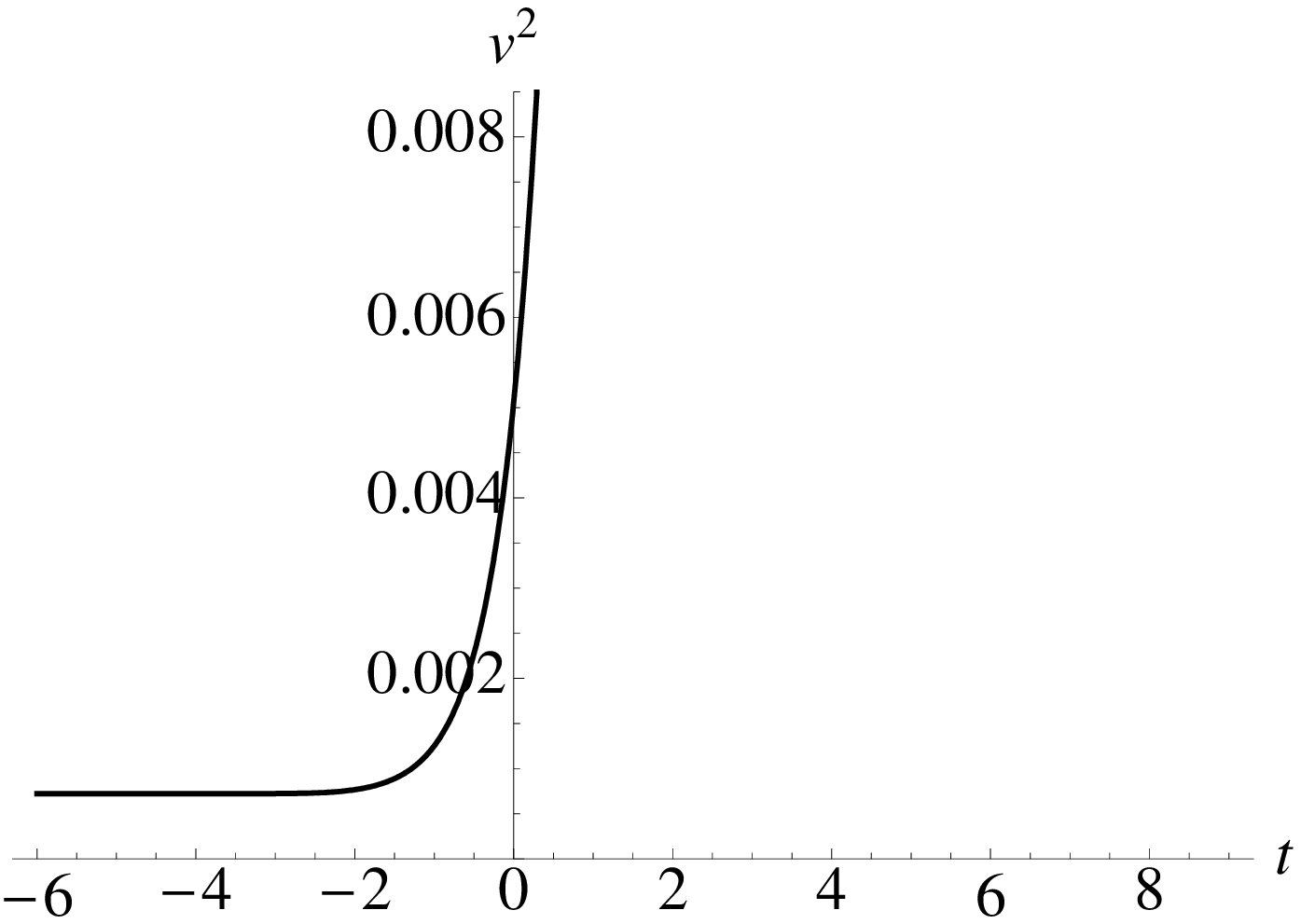}
 \includegraphics[width=0.35\textwidth]{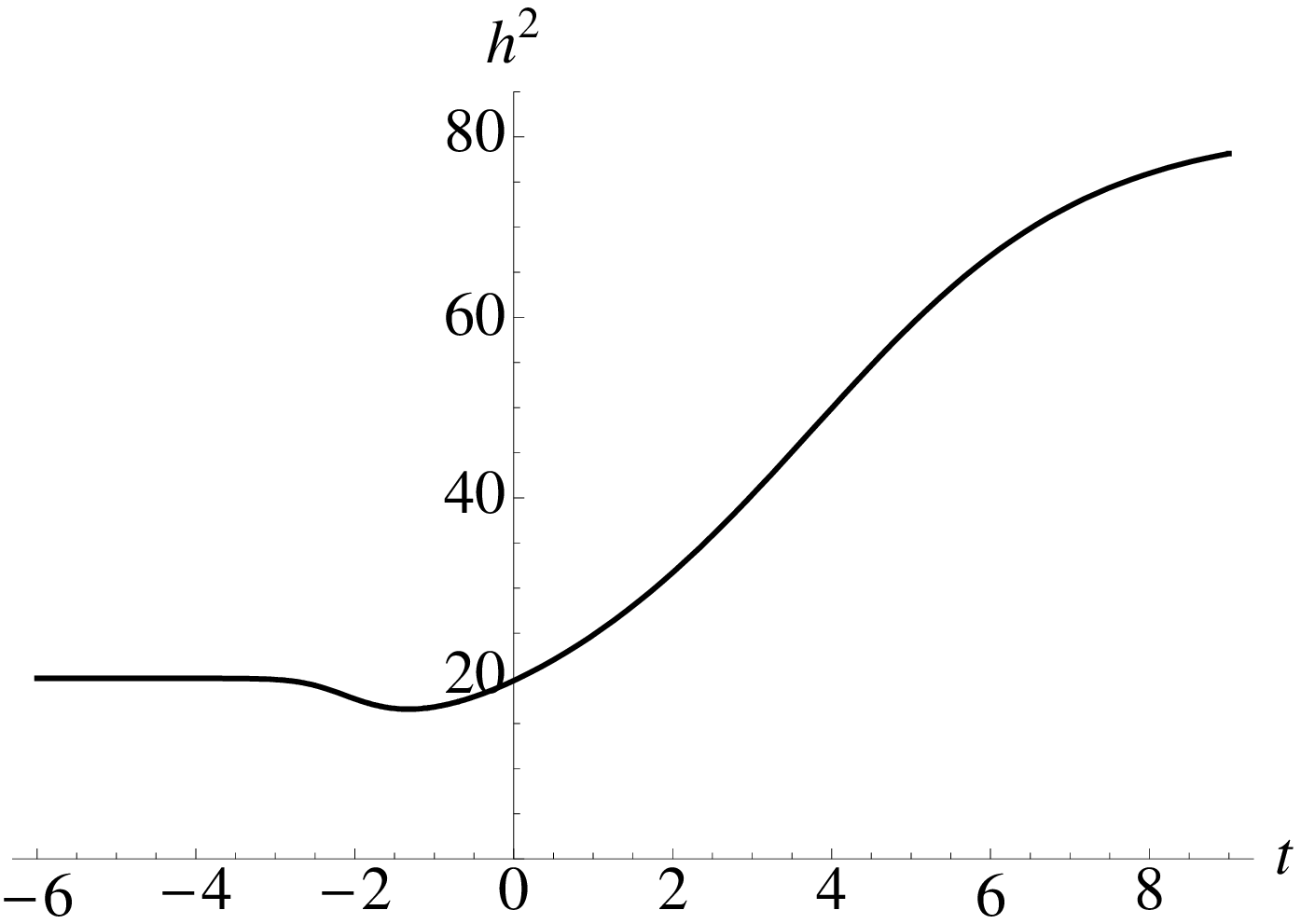}
 \includegraphics[width=0.35\textwidth]{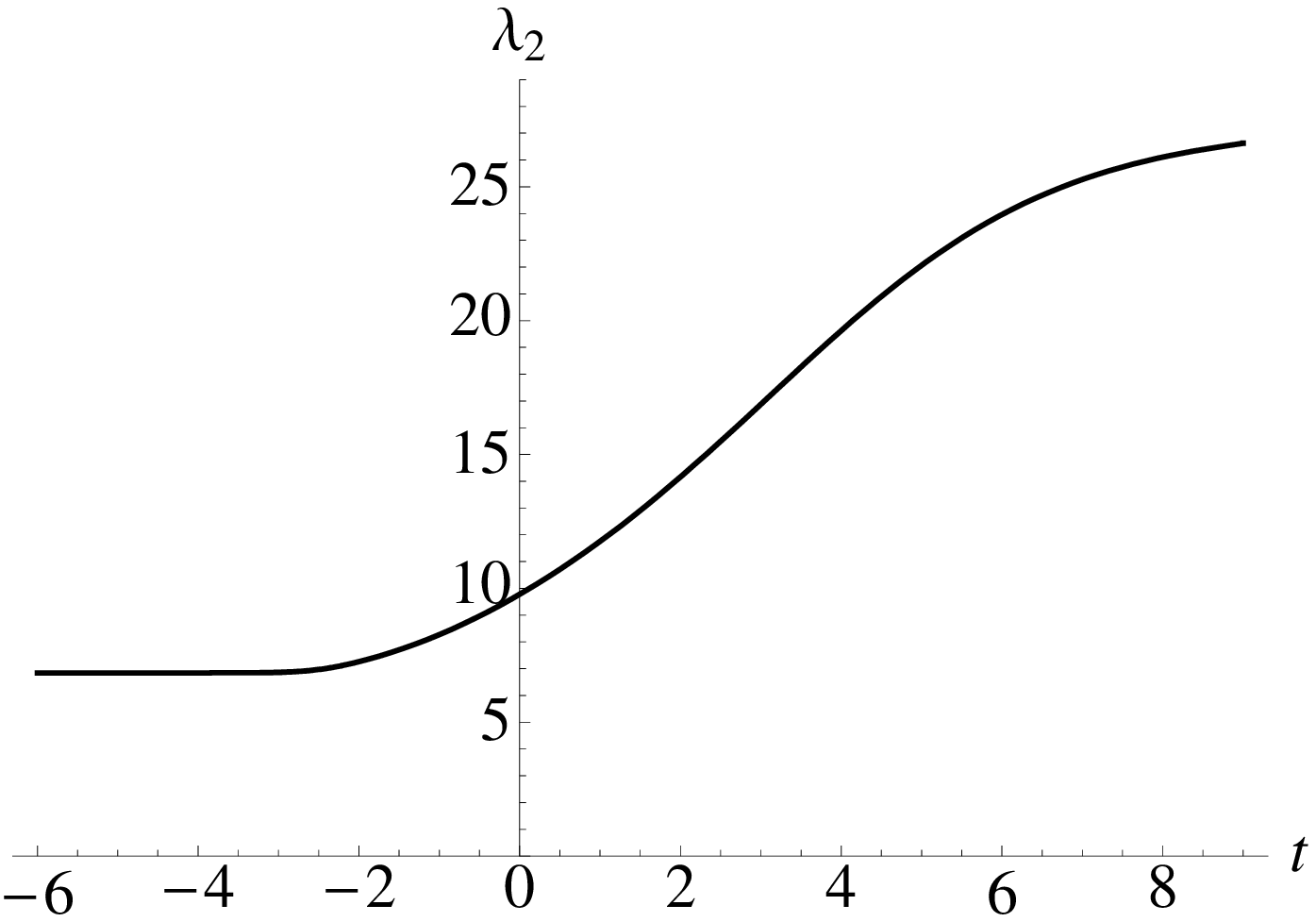}
 \caption{Typical flows from the non-Gau\ss ian UV fixed point to the IR. The
   upper plot displays the running of $2 \kappa k^2=v^2$. It runs to a
   constant value in the IR, corresponding to the properties of the SSB
   phase. The middle and the lower plot show the running of $h^2$ and
   $\lambda_2$ from the UV fixed point regime via the crossover regime to the
   IR decoupling regime as described in the text.}
\label{fig:ssbflow}
\end{center}
\end{figure}

Of course, the present theory with a small number of fermion flavors cannot be
expected to resemble the standard model also quantitatively. For instance, for
the $\Nf=\frac{1}{10}$ model in the leading-order truncation, we have not
found an RG trajectory connecting the non-Gau\ss ian fixed point with an IR
limit where the Yukawa coupling approaches its standard-model value $h_{k\to
  0} =m_{\text{top}}/v \simeq 0.711$. As a simple example, we have studied a
universe with a top mass given by $m_{\text{top}} = h v|_{k\to 0}$ with
$h_{k\to0} = 20.0$. Our prediction for the Higgs mass then is
$m_{\text{Higgs}} = \sqrt{6.845} v$. We have explicitly checked that
$m_{\text{Higgs}}$ indeed does not depend on the starting point in the
fixed-point regime nor on the UV cutoff scale $\Lambda_{\text{UV}}$. In
Fig.~\ref{fig:ssbflow}, a generic flow of the three running couplings of the
$\phi^4$ truncation is depicted.  We have also tested the system for $\Nf=1/15$
and with a top mass corresponding to $h_{k\to 0}=7.0$ which gives
$m_{\text{Higgs}}=\sqrt{2.9} v$.

The observations made here in the leading-order truncation also hold at
next-to-leading order. An interesting property of the asymptotically safe
system is that the set of possible IR values for the renormalized couplings is
generically bounded, i.e., not all conceivable values of the physical
parameters lie on RG trajectories that emanate from the non-Gau\ss ian fixed
point. In the present case, we observe that it is not possible to tune the flow for
$h_{k\to0}$ below a certain value $h_{\text{min}}$, implying that a lower
bound on $m_{\text{top}}/v$ exists.

The mechanism behind this property can easily be understood. The RG flow
passes essentially through three regimes: the fixed-point regime in the UV,
the decoupling regime in the IR, and a crossover regime in-between. In the UV
fixed-point regime, the system behaves conformally and all couplings stay
close to their fixed point values. The decoupling regime in the IR is
characterized by the freeze-out of the dimensionful Higgs vacuum expectation
value $v$. This generates Higgs and top masses which decouple from the flow at
scales $k\ll v$. No massless modes remain and all running couplings freeze out
and approach their IR values. The crossover regime in-between connects the UV
fixed-point with the IR decoupling regime. Here, all couplings run fast and
the full nonperturbative dynamics becomes important. Generically all $\beta$
functions are sizeable here, so that the system spends little ``RG time'' in
the crossover regime. This implies that the Yukawa coupling has little RG time
to run from its large UV fixed-point value to its physical IR value, before it
freezes out due to decoupling. 

Technically, we observe that there is a very robust IR fixed point in the SSB
flow equations that attracts the flow towards the IR in the crossover
regime. In the simple $\phi^4$ truncation, it is characterized by fixed-point
values $\kappa_{\text{IR}}^\ast,h_{\text{IR}}^{\ast 2},
\lambda_{2,\text{IR}}^\ast$.  This fixed point is not directly physically relevant
as $\kappa_{\text{IR}}^\ast<0$. The SSB constraint $\kappa>0$ however implies
that the physical Yukawa coupling is bounded by this IR fixed-point value,
$h_{k\to 0}> h_{\text{IR}}^\ast$. 
We conclude that an asymptotically safe Yukawa system offers a natural
explanation for a large top mass.

\section{Conclusions}
\label{sec:conc}

We have used a simple $Z_2$-invariant Yukawa system with massless fermions and
a real scalar field in order to explore the possibility of asymptotically safe
Yukawa systems for a solution of the triviality and hierarchy problem in the
standard model. Employing the functional RG within a systematic and consistent
derivative expansion of the effective action, we have identified a new
mechanism based on a conformal Higgs vacuum expectation value that can give
rise to a non-Gau\ss ian interacting stable UV fixed point. By contrast, our
study of the symmetric regime at vanishing expectation value confirms the
triviality of the Higgs sector nonperturbatively, as both essential couplings,
the Yukawa coupling and the $\phi^4$ interaction, are marginally irrelevant. 

Whether or not asymptotic safety based on a conformal Higgs expectation value
is realized, depends on the algebraic structure of the theory, in particular
on the relative strength between bosonic and fermionic fluctuations. Our
results show that this Yukawa model is not asymptotically safe for $\Nf=1$ or
larger, since fermionic fluctuations dominate the running of the expectation
value, rapidly driving the system to decoupling of massive modes. Allowing for
arbitrary positive real values of $\Nf$, we discover Yukawa systems at
$\Nf\lesssim 0.3$ where a conformal expectation value induces the desired
non-Gau\ss ian fixed point and renders the models asymptotically safe. 

These models do not have a triviality problem but can be extended to
arbitrarily large momentum scales and are thus candidates for a fundamental
quantum field theory. Within the derivative expansion, we find evidence that
the hierarchy problem can be less severe in these systems in the sense that the
largest relevant coupling renormalizes softer than quadratically. 

There are two most attractive features of such asymptotically safe Yukawa
systems. First, the UV fixed point exhibits only two RG relevant directions, implying
that the IR physics of the system is fully determined by fixing two physical
parameters. For instance, once the top mass and the Higgs vacuum expectation
value are fixed, the Higgs mass is a true and parameter-free prediction of the
model. This clearly distinguishes our scenario from many other suggestions of
physics beyond the standard model where the hierarchy and triviality problem
are solved or reparameterized at the expense of an increasing number of
parameters. Second, the range of possible physical IR parameters is
generically bounded, since the set of possible RG trajectories emanating from
the UV fixed-point map out a bounded IR parameter region. In the present
example, the top mass is bounded from below for models in the Higgs phase of
spontaneous symmetry breaking. Therefore, our scenario offers a natural
explanation for large values of the top mass. 

Even though our basic concept does deliberately not invoke supersymmetric
structures, there are certain similarities to properties of supersymmetric
theories. In the latter, supersymmetric partners to standard-model particles
are introduced in order to enhance the symmetry of the theory. This enhanced
symmetry guarantees substantial cancellations between bosonic and fermionic
fluctuations, such that crucial parameters such as the Higgs mass remain
protected against strong renormalization. In our case, there is no enhanced
symmetry, but the dynamical properties of the theory itself make it possible
that fermionic and bosonic fluctuation contributions to the vacuum expectation
value $v$ balance each other and induce a conformal threshold behavior over
many orders of magnitude. 

From a technical perspective, asymptotically safe scenarios are difficult to
establish due to the inherent nonperturbative and hence less controllable
nature. This is particularly true if a possible non-Gau\ss ian UV fixed point
is a pure strong-coupling phenomenon. In the present case, asymptotic safety
does not predominantly rely on strong coupling but is induced by the conformal
threshold behavior of the Higgs expectation value. Even though threshold
physics is also inherently nonperturbative, it does not require particularly
strong coupling. Our findings indicate, that the derivative expansion of the
effective action is not only systematic and consistent, but also well suited
to reliably explore this asymptotic-safety scenario: the anomalous dimensions,
serving as effective expansion parameters, remain small, and all qualitative
features are robust when going from leading-order to next-to-leading order. On
the other hand, reliable quantitative results for physical quantities such as
the UV critical exponents or the IR Higgs mass prediction certainly require a
larger truncation particularly in the boson-fermion coupling sector. 

In the light of our results, a search for asymptotically safe Yukawa systems
also by other methods would be highly welcome. For the present simple system,
small-$\Nf$ expansion techniques appear to be most promising. These indeed
exist within the worldline approach to quantum field theory, where subclasses
of infinitely many Feynman diagrams can be resummed into closed worldline
expressions order by order in a small-$\Nf$ expansion. These methods have been
shown to work in pure scalar models so far \cite{Gies:2005sb} and need to be
generalized to fermions. Of course, also many non-perturbative lattice studies
of Yukawa systems exist, see \cite{Smit:1989tz} for reviews, recently also
employing chirally-invariant lattice fermions \cite{Gerhold:2007gx}. The fact
that no non-Gau\ss ian fixed points have been found in standard settings so
far (for alternative higher-derivative models, see \cite{Jansen:1993jj}) is
well in accord with our findings. Since lattice realizations often have a
large fermion number, the resulting fermion dominance generically inhibits a
conformal threshold behavior required for our scenario. Moreover, since our
non-Gau\ss ian fixed point has two relevant directions, a tuning of two bare
parameters is required in order to take the corresponding continuum limit on a
line of constant physics. Even though this is certainly possible on the
lattice, this point can easily be missed during a general lattice study.

Whereas Yukawa systems with  $\Nf\lesssim 0.3$ are unlikely to be relevant for
the true Higgs sector of the standard model, our results clearly indicate that
an asymptotic safety scenario based on a conformal threshold behavior requires
a balancing between bosons and fermions. Hence, we expect that a variety of
realistic models exists where the fermion contributions are balanced by a
sufficiently large bosonic sector. Work in this direction also involving
chiral structures is in progress. 

\acknowledgments

The authors are grateful to S.~Fl\"orchinger, C.~Gneiting, R.~Percacci and
C.~Wetterich for interesting discussions and helpful comments. This work was
supported by the DFG under contract No. Gi 328/1-4 (Emmy-Noether program), Gi
328/5-1 (Heisenberg program) and FOR 723.

\end{document}